\documentclass{article}

\usepackage{microtype}
\usepackage{graphicx}
\usepackage{subfigure}
\usepackage{booktabs} 

\usepackage{hyperref}

\usepackage[accepted]{icml2025}

\usepackage{amsmath}
\usepackage{amssymb}
\usepackage{mathtools}
\usepackage{amsthm}

\usepackage[capitalize,noabbrev]{cleveref}

\theoremstyle{plain}

\theoremstyle{definition}

\theoremstyle{remark}

\usepackage[disable,textsize=tiny]{todonotes}

\icmltitlerunning{Deprecating Benchmarks}

\begin{document}

\twocolumn[
\icmltitle{Deprecating Benchmarks: Criteria and Framework}


\icmlsetsymbol{equal}{*}

\begin{icmlauthorlist}
\icmlauthor{Ayrton San Joaquin*}{AISL}
\icmlauthor{Rokas Gipiškis*}{AISL,Vil}
\icmlauthor{Leon Staufer*}{tum,lmu}
\icmlauthor{Ariel Gil}{AISL,tl}

\end{icmlauthorlist}

\icmlaffiliation{AISL}{AI Standards Lab, Dover, Delaware, United States}
\icmlaffiliation{tum}{Technical University of Munich, Munich, Germany}
\icmlaffiliation{lmu}{LMU Munich, Munich, Germany}
\icmlaffiliation{Vil}{Institute of Data Science and Digital Technologies, Vilnius University, Vilnius, Lithuania}
\icmlaffiliation{tl}{Trajectory Labs, Toronto, Canada}

\icmlcorrespondingauthor{Rokas Gipiškis}{rokas.gipiskis@mif.vu.lt}

\icmlkeywords{Machine Learning, Benchmarks, Evaluations, Datasets, Safety, Maintenance, Frontier Models}

\vskip 0.3in
]



\printAffiliationsAndNotice{*Core contribution} 

\begin{abstract}
As frontier artificial intelligence (AI) models rapidly advance, benchmarks are integral to comparing different models and measuring their progress in different task-specific domains. However, there is a lack of guidance on \emph{when} and \emph{how} benchmarks should be deprecated once they cease to effectively perform their purpose. This risks benchmark scores over-valuing model capabilities, or worse, obscuring capabilities and safety-washing. Based on a review of benchmarking practices, we propose criteria to decide when to fully or partially deprecate benchmarks, and a framework for deprecating benchmarks. Our work aims to advance the state of benchmarking towards rigorous and quality evaluations, especially for frontier models, and our recommendations are aimed to benefit benchmark developers, benchmark users, AI governance actors (across governments, academia, and industry panels), and policy makers.

\end{abstract}

\section{Introduction}
\label{intro}
Benchmarks are fundamental to evaluating model outputs directly and remain the dominant form of evaluating artificial intelligence (AI) systems over time. 
Within AI risk assessment and mitigation, benchmarks serve as critical indicators for when frontier models\footnotemark exhibit dangerous capabilities. They are becoming standard tools mandated by various regulations worldwide, notably the European Union's AI Act (Article 55 in particular) \citep{aiact} and China's proposed AI law \citep{linghan2024chinagov}.

\footnotetext{We use the definition of frontier models from \citet{anderljung2023frontier}: ``highly capable foundation models that could exhibit sufficiently dangerous capabilities.''}

Yet, many benchmarks in use today are outdated, flawed, or misaligned with their intended purpose. Some persist simply because they are historically tied to influential models, thereby functioning as de facto standards despite their limitations (e.g. ImageNet with AlexNet, \citealt{beyer2020we}). This prevents better alternatives from gaining traction (e.g. the Benchmark Lottery introduced by \citealt{dehghani2021benchmark}). 

Commercial incentives further entrench this problem. AI labs have little incentive to deprecate benchmarks that support the superiority of their models. Flawed or outdated benchmarks often inflate model performance, enabling easy claims of state-of-the-art results without meaningful progress \citep{eriksson2025can}. A notable example is Meta's LLaMA 4, which was fine-tuned specifically for conversational benchmarks \citep{robisonMetaGotCaught2025}. Changing benchmarks risks disrupting leaderboard continuity, invalidating prior claims, and complicating public messaging \citep{campoloStateoftheartTemporalOrder2025}. Furthermore, benchmark developers may resist deprecation to protect prior work, internal tooling, and reputational capital \citep{eriksson2025can}. With no regulatory mandate to reassess benchmark validity, the market defaults to inertia, perpetuating benchmarks that no longer validly measure model capabilities and are broadly applied to every context \citep{raji2021ai}.

We argue that outdated or flawed benchmarks must be actively deprecated to prevent distorted capability assessments, wasted resources on ineffective evaluations \citep{varoquaux2024hype, polo2024tinybenchmarks}, and safety-washing \citep{ren2024safetywashing}. By deprecation, we mean to avoid the use of the target benchmarks for evaluation of current and future models. We introduce criteria and a process to deprecate or partially deprecate (i.e., update) inadequate benchmarks. While ideally this process involves both benchmark creators and governance actors (spanning governments, academia, and industry panels), it can also be initiated independently by governance actors when creators are unavailable. Our proposal increases transparency in benchmark evaluations and complements broader governance efforts across the benchmark development and usage lifecycle.

\section{Review of the Benchmarking Literature}
\label{review}

\textbf{Limitations of Current Benchmarks}
Numerous studies have highlighted problems in popular benchmarks or benchmarking practices more broadly \cite{blagec2023benchmark, mcintosh2024inadequacies, ren2024safetywashing, singh2025leaderboard, dehghani2021benchmark, anwar2024foundational}. \citet{reuel2024betterbenchassessingaibenchmarks} analyzed 24 benchmarks, mainly for foundation models, and found that even common benchmarks have significant flaws. \citet{liao2021we} provides a taxonomy of benchmark failure modes across multiple domains. \citet{rauh2024gapsinsafetyeval} highlighted the flaws of current benchmarking practices by evaluating model capabilities separated from its context, especially with automated benchmarking.
\citet{eriksson2025can} showed that current benchmarks are fragile tools for risk assessment and mitigation. Importantly, they highlighted the problem of determining which benchmarks to trust. Our paper attempts to  answer this via its negation: deprecation implies that a benchmark should not be trusted for a particular context.

\textbf{Qualities of Good Benchmarks}
Numerous studies have worked on improving benchmark quality in other fields, such as software engineering \cite{miltenberger2023benchmarking, beyer2019reliable}.

For AI, \citet{reuel2024betterbenchassessingaibenchmarks} reviewed benchmarking best practices from other fields and developed criteria to assess benchmarks targeting different parts of its lifecycle. \citet{alonso2025evaluating} introduced a set of principles for constructing quality benchmarks.
Similarly, \citet{davis2023benchmarks} lists the purposes, requirements, and characteristics of desirable commonsense reasoning benchmarks for AI. It acknowledged the need for model developers to scrutinize the benchmarks they use, and either abstain, fix, or report flawed benchmarks they come across.

\textbf{Dataset Maintenance and Deprecation}
Version control in the form of a Direct Object Identifier (DOI) \cite{paskin2010digital} has been proposed for datasets to track changes and aid in the deprecation of them \citep{luccioni2022framework, ghosh2025ailuminate}. Meanwhile, \citet{stauferAuditCardsContextualizing2025} argued that evaluations such as benchmarks should include information about the circumstances under which they become deprecated.

Our work is closest to, and takes inspiration from \citet{luccioni2022framework}, which introduced the ideas of a deprecation framework, reporting, and a central repository for deprecated datasets. We build on their findings in light of recent trends with the rise of frontier models. We provide guidance to governance actors on their role in aiding developers in the deprecation process towards a robust evaluation ecosystem. Notably, we address the possibility of partial deprecation (which we also refer to as \emph{upgrading}) and provide criteria specific to benchmarks.

Furthermore, our focus on benchmarks presents easier deprecation compared to datasets in general. For frontier models, benchmarks have transformed into mainly test-time artifacts, where there is no need to explicitly train the model, either via pretraining or fine-tuning, to complete the benchmark task. This opens the avenue for benchmarks to have a cheaper and faster deprecation process because they can be immediately applied to the models that the previous benchmark was applied to, unlike training datasets. This has implications for governance, which we discuss in our recommendations.

\section{Criteria for Deprecation}
\label{criteria}

Based on our review of benchmarking practices in \Cref{review}, we present a non-exhaustive list of criteria for benchmark deprecation. For each criterion, we provide a short description and, where applicable, an illustrative example.\footnote{These examples are intended solely for illustrative purposes, and we do not claim that benchmarks should be deprecated in all instances, as determining concrete thresholds for these criteria is beyond the scope of this paper.} 

The criteria could be grouped into two broad categories: (1) quantitative signals of diminishing utility, such as benchmark saturation, memorization or statistical bias, and (2) qualitative issues stemming from fundamental design flaws or shifts in the broader context of use. These include annotation errors, task obsolescence, invalidated assumptions, and the evolving relevance of the benchmark’s task. For each criterion, we provide a short description and, where applicable, an illustrative example.

We emphasize that \textbf{the list is intentionally non-exhaustive}. Rather than providing rigid, binary rules, the criteria are meant to act as a flexible and evolving set of heuristics. They should be treated more like “soft” guidance, similar to case law, where precedent, accumulated experience, and community norms inform which benchmarks should be deprecated over time. We expect that new criteria will emerge through ongoing dialogue across the AI evaluation community, including governance panels, benchmarking workshops, and participatory audits.

\begin{enumerate}
    \item \textit{Saturation} Benchmark saturation refers to benchmarks reaching or approaching their evaluation ceiling, such that further improvements provide limited information about a model's true capabilities. This has been observed \cite{barbosa2022mapping, maslej2025aiindex} in various benchmarks, such as MMLU \cite{hendrycks2020measuring}, GSM8K \cite{cobbe2021training}, and HumanEval \cite{chen2021evaluating}. When benchmarks are close to saturation, they become less effective measures, where more nuanced capability gains might not be detected.

    \item \textit{Contamination} When models memorize benchmark data due to leakage or contamination\footnote{This paper remains neutral on the usefulness of fully public benchmarks or benchmarks without private holdout sets and considers it out of scope.}, their performance no longer reflects true generalization with respect to the measured capabilities. This undermines the benchmark’s ability to accurately assess those capabilities. Various instances of benchmark leakage have been observed \cite{zhou2023don, xu2024benchmark, xu2024benchmarking, ravaut2024much}.
    
    \item \textit{Statistical bias} Poorly balanced classes or label distributions can skew benchmark results, leading models to exploit dataset imbalances rather than demonstrating the intended capabilities. For example, a model may perform well on an anomaly-detection benchmark simply by answering correctly on the majority benign samples, even though it fails to detect the minority anomalous samples.

    \item \textit{High annotation error rate} Errors introduced by benchmark creators or labelers/annotators, whether human or automated, can compromise the quality of benchmark data. A high error rate leads to unreliable benchmarks and inaccurate model evaluations. For example, \citet{gema2024we} found that the Virology subset of the Massive Multitask Language Understanding (MMLU) benchmark \cite{hendrycks2020measuring} contained errors in 57\% of the analysed questions, and that 6.49\% of all MMLU questions had errors.

    \item \textit{Task obsolescence} A benchmark may become obsolete if the underlying task is no longer considered relevant or has effectively been solved. Continued use of such a benchmark, due to familiarity or research inertia, can lead to misleading signals of progress and wasted resources. While related to saturation, this criterion focuses on the evolving relevance of the task itself rather than model performance limits. For example, the BIG benchmark\cite{srivastava2022beyond} weighs the task of generating words given a bag of letters equally as that of causal reasoning. The former task may be obsolete for developing reasoning models.
    
    
    \item \textit{Invalidated assumptions} Benchmarks often rely on simplifying assumptions that may become inappropriate as the field evolves. When such assumptions no longer hold, the benchmark fails to reflect real-world complexity or task diversity. For instance, the commonly used Needle-in-a-Haystack test \cite{kamradt2024needle} for evaluating in-context retrieval in long context LLMs may not accurately represent many real-world retrieval-augmented generation (RAG) applications. Unlike the test's focus on retrieving a single fact, RAG typically involves retrieving multiple pieces of information and reasoning over them \cite{langchain2024multi}.
    
    \item \textit{Semantic drift} Over time, the meaning or interpretation of a task or its labels can change, just like how the meanings of words or cultural associations shift over time \cite{kutuzov2018diachronic}. For example, \citet{luccioni2022framework} describe datasets as static or frozen in time at the moment of their creation. As language, context, or social conventions evolve, benchmarks can become outdated or unrepresentative.
\end{enumerate}

\section{Deprecation Framework}
\label{framework}

The deprecation framework consists of three phases: \textit{assessment}, \textit{reporting}, and \textit{notification}, with reporting including a deprecation report that provides key details for benchmark users. Deprecation levels range from \textit{updating} (which preserves valid components while addressing problematic elements like annotation errors, copyrighted material, or outdated task definitions) to \textit{complete deprecation}.

\subsection*{Deprecation by Governance Actors or Other Entities}
Our framework permits benchmark deprecation by governance actors, such as government agencies or industry panels. Benchmark developers possess crucial knowledge of its properties, naturally positioning them to identify when deprecation becomes necessary. However, some deprecation decisions (e.g. invalidated assumptions from \Cref{criteria}) are better made by external stakeholders such as users and researchers \citep{rajiAIEverythingWhole2021}. Furthermore, some benchmarks are abandoned and there are no developers to issue a deprecation notice \citep{luccioni2022framework}. 

In such cases, third-party deprecation by governance actors---including government agencies, industry consortia, and public-interest groups---using the same framework is essential. These actors can create \emph{deprecation lists} of curated deprecated benchmarks alongside deprecation reports, though inclusion requires additional transparency given differing motivations. We detail specific recommendations in \Cref{recommendations} and describe example scenarios in \Cref{appendix:cases}.

\subsection{Assessment}
\label{framework:assessment}

Assessment determines whether deprecation is necessary by evaluating potential impacts and examining which benchmark components remain valid via the criteria in \Cref{criteria}. This process should be performed regularly, including tracking performance saturation curves, examining emerging literature critique, and soliciting feedback from communities who use or helped curate benchmark data.

Governance actors should establish a formalised appeals process, allowing benchmark developers or other parties to contest deprecation decisions. The process should at minimum provide a contact person, a timeline to resolution, and enable proposals for partial deprecations as alternatives to full deprecation. A review board of independent experts should be integrated to increase accountability, with affected communities included in the assessment and consultations with original benchmark developers when third-party actors are involved.

The deprecation level (partial versus full) depends on the nature and severity of identified issues. Benchmarks with statistical bias or contamination may suit partial deprecation through corrective measures such as resampling to rebalance class distributions, incorporating random variables, or removing leaked data. In contrast, task obsolescence or invalidated assumptions typically require full deprecation, particularly when fundamental misalignment exists between the benchmark and current use contexts—for instance, benchmarks centred on irrelevant tasks (e.g. word unscrambling for reasoning evaluation) or outdated design principles (e.g. single-fact retrieval in complex RAG scenarios). For annotation errors, deprecation level depends on their prevalence and impact: minor, correctable labelling issues may warrant partial deprecation with documented corrections, while substantial portions containing flawed data may require full deprecation of the affected subset or entire benchmark.

\subsection{Reporting}
\label{framework:reporting}
In the second reporting step, the deprecation report should contain clear rationale and underlying risks that necessitate the deprecation decision, supported by evidence from the assessment phase. The report should provide explicit instructions regarding future benchmark usage, distinguishing between full deprecation and partial deprecation of specific tasks or subcomponents. For upgrading, the report should clearly delineate which benchmark components remain valid and address how past benchmark usage should be interpreted given the changes.

The reports should outline implementation timelines with strategies to mitigate potential harms from continued benchmark use, identifying an updated version or alternative benchmarks to facilitate transition if any exist. It should also include a process to allow authorized use of the deprecated benchmark, such as for research and archiving purposes \citep{luccioni2022framework}. For a deprecation report made by a governance actor, it should also include an appeals process described in \Cref{framework:assessment}.

The report should also include information on the actors performing the deprecation with their contact information, details on the benchmark, and its developers. This information is important for contextual transparency \citep{gebruDatasheetsDatasets2021}.

To illustrate the structure and content of a deprecation report, we provide two \textit{partially fictional} case studies based on the SWE-Bench benchmark shown in \Cref{appendix:deprecation_report}. The first simulates a partial deprecation of SWE-Bench v1.0 due to methodological issues, while the second shows a partial deprecation  of the SWE-Bench Lite v1.0, focusing on task subsets with ambiguous specifications and unreliable test coverage. A full deprecation process is very similar to the partial one presented here.

\subsection{Notification}
\label{framework:notification}
The final step of deprecation is notification and prevents continued use of deprecated benchmarks and reduces disruption to the evaluation ecosystem. Deprecation notices should preferably appear in the same channels as the original benchmark publication. Clear visual indicators or metadata should distinguish deprecated benchmarks from active ones when listed in catalogs or referenced in papers, functioning similarly to academic retraction notices \citep{barbourGuidelinesRetractingArticles2009}. Key benchmark users whose evaluations may be invalidated (e.g. in system-critical settings) will benefit from receiving direct notification of the deprecation. For upgrading, version control should be used to differentiate between original and modified benchmark versions.

\section{Adapting the Framework into Practice}
\label{adapting}
We provide a detailed blueprint for adapting deprecation in the European Union (EU), where the AI Office (AIO) could create deprecation lists of benchmarks related to safety-critical and dual-use capabilities such as CBRN capabilities \cite{barrett2024benchmark}. Under the European Commission, AIO is mandated by the EU AI Act to manage high-risk model evaluations (Article 15.2) and supplement benchmarks for classifying general-purpose AI models with systemic risk (Article 51.3) \cite{aiact}. Providing deprecation lists exemplifies both managing evaluations and supplementing benchmarks, with AIO able to accredit third parties for less-immediate risks or specialised contexts.

\textbf{Assessment} AIO compiles and periodically reviews benchmarks commonly used for safety-critical tasks (e.g. biological capabilities), determining which require full or partial deprecation. 
Decisions are informed by inspecting benchmark scores in model cards, weighing research showing contamination or criticism, directly testing models, or convening expert consultations based on given criteria. 
This phase produces a deprecation list containing benchmarks for full or partial deprecation.

\textbf{Reporting} AIO creates a deprecation report including, for each benchmark: (i) full or partial deprecation decision and affected version, (ii) deprecation rationale, (iii) timeline for the deprecation process, (iv) mitigation details including alternative benchmarks or upgrade specifications, and (v) directions for interpreting past results. The report should permit authorised use for meta-research or legal investigations and establish an appeals process for developers to contest decisions, particularly when full deprecation might be reduced to partial. A full example appears in \Cref{appendix:deprecation_report}.

The report should allow authorized use of the deprecated benchmarks, including for meta-research on benchmarks or legal investigations. The AIO could set up an appeal process for developers to contest deprecation, especially in the case when a full deprecation can be changed to a partial one.

\textbf{Notification} AIO contacts EU member states' national AI authorities and affected AI programmes, either via periodic notifications after scheduled maintenance or immediate notices following critical vulnerabilities. Models deployed commercially in the EU must update model cards and technical reports after notification within a timeframe specified by the AIO. Listing benchmark versions should be standard practice to enable user scrutiny when immediate updates are not realised, allowing users to independently verify model compliance with the latest deprecation list.


\section{Recommendations}
\label{recommendations}
We reiterate our key insights as recommendations to benchmark developers, policymakers, and governance actors:
\begin{enumerate}
    \item \textbf{Establish version control for benchmarks.} Following \citet{luccioni2022framework, ghosh2025ailuminate}, benchmarks require version control to determine when changes are made and to avoid invalid use and comparisons. The version should be immediately visible alongside the benchmark name in all references \citep{reuel2024betterbenchassessingaibenchmarks} and point towards the updated version, especially in deprecation lists. We recommend using the standard versioning system for digital artifacts: the Digital Object Identifier (DOI) \citep{paskin2010digital}. Versioning encourages model providers to remain current with benchmarking their model. For example, all models that report being tested before the date of benchmark version change are assumed to have been tested under the previous version of the benchmark. (Supports \emph{Notification} \ref{framework:notification})
\end{enumerate}

Benchmark developers are essential to deprecation because they are most familiar with the benchmark. We recommend the following, as introduced by \citet{luccioni2022framework}:

\begin{enumerate}
  \setcounter{enumi}{1}
    \item \textbf{Have a deprecation plan when creating a benchmark.} This includes, at a minimum, a plan for archiving benchmark-related artifacts (e.g. data and code to generate the dataset). (Supports \ref{framework:assessment} \emph{Assessment})

    \item \textbf{Establish a point of contact.} They should handle queries related to the deprecated benchmark, including access to it for authorized use. (Supports \ref{framework:notification} \emph{Notification})
\end{enumerate}

We believe external governance parties (e.g. government AI authorities, scientific panels, industry consortia) are essential to the enforcement of deprecation and encouraging a healthy evaluation ecosystem. Therefore, we recommend the following:

\begin{enumerate}
  \setcounter{enumi}{3}
    \item \textbf{Create context-specific deprecation lists.} Since benchmarks are not neutral \citep{rauh2024gapsinsafetyeval} nor universal \citep{raji2021ai}, creating deprecation lists should depend on the context of the evaluation, factoring in both the socio-technical context of the model's intended use and the competency of evaluating parties. For instance, internal development teams may focus primarily on safety benchmarks, while public model releases require broader evaluation across diverse deployment scenarios. Deprecation lists creation should take this context of the evaluation into account.

\item \textbf{Construct deprecation lists via transparent practices.} They must have verifiable claims for deprecation and informed thorough analysis of the issues that the benchmark in question has. (Supports \ref{framework:assessment} \emph{Assessment})

\item\textbf{For every benchmark, clearly communicate the reason for including it in the deprecation list.} This is fundamental in maintaining trust in the deprecation process, especially when it is not done by the benchmark developers themselves, (Supports \ref{framework:reporting} \emph{Reporting})

\item\textbf{As a compliance requirement for model developers, discourage the use of deprecated benchmarks.} Preventing models from being accessed because they are tested on deprecated benchmarks can accelerate the retirement of problematic benchmarks. For research models, this can be facilitated by conference venues \citep{luccioni2022framework}. For internal and commercial models, this can be managed by both industrial consortia and government agencies (e.g. a safety institute). For models designed to serve a particular community, this opens a new avenue for community input on the evaluation process. (Supports \ref{framework:notification} \emph{Notification})

\item\textbf{Establish an appeal mechanism for deprecation lists.} This promotes trust by ensuring that the deprecation list remains responsive to feedback from benchmark developers, especially in cases where the inclusion criteria may be flawed.  (Supports \ref{framework:assessment} \emph{Assessment})

\end{enumerate}

\clearpage
\section*{Impact Statement}

This paper presents work whose goal is to advance the field of Machine Learning, in particular for better measurement of model capabilities and safety relevant properties. Benchmark deprecation is discussed as a potential solution, but it may not be sufficient on its own to address the problems highlighted in this paper. 

\section*{Contribution Statement}
ASJ: Conceptualized the topic of the paper. Wrote \Cref{intro}, \Cref{review}, \Cref{adapting}, \Cref{recommendations}, \Cref{appendix:cases} and part of \Cref{framework}. Led the project administration. Reviewed, revised, and copy edited all sections.

RG: Developed the criteria for deprecation and contributed to the early conceptualization of benchmark deprecation. Contributed to the initial literature review. Wrote \Cref{criteria} and part of \Cref{framework:assessment}. Reviewed, revised, and copy edited all sections.

LS: Contributed to the early conceptualization of benchmark deprecation, especially the deprecation framework. Wrote \Cref{framework}, \Cref{appendix:deprecation_report}, and parts of \Cref{intro}.  Reviewed, revised, and copy edited all sections.

AG: Gave feedback on early paper drafts and helped narrow topics. Wrote an early draft of \Cref{recommendations}. Reviewed, revised, and copy edited various sections - primarily \ref{framework}, \ref{criteria}, and \ref{intro}. Wrote the template section (with help from LS), adding real world examples of SWEbench \Cref{appendix:deprecation_report}.


\section*{Acknowledgements}
We thank Koen Holtman for providing comments to the content of the paper.

The AI Standards Lab, where this work was primarily performed,  is funded by the AI Safety tactical opportunities fund (grant number A2003126), funding from Open Philanthropy (\href{https://www.openphilanthropy.org/grants/ai-standards-lab-ai-standards-and-risk-management-frameworks/}{https://www.openphilanthropy.org/grants/ai-standards-lab-ai-standards-and-risk-management-frameworks/}), and funding from the Survival and Flourishing Fund (\href{https://survivalandflourishing.fund/}{https://survivalandflourishing.fund/}). 

Leon Staufer was supported by the \href{https://www.matsprogram.org/}{ML Alignment Theory Scholars (MATS) program}.



\bibliography{icml_db}

\begin{thebibliography}{47}
\providecommand{\natexlab}[1]{#1}
\providecommand{\url}[1]{\texttt{#1}}
\expandafter\ifx\csname urlstyle\endcsname\relax
  \providecommand{\doi}[1]{doi: #1}\else
  \providecommand{\doi}{doi: \begingroup \urlstyle{rm}\Url}\fi

\bibitem[Alonso \& Church(2025)Alonso and Church]{alonso2025evaluating}
Alonso, O. and Church, K.
\newblock {Evaluating the Evaluations: A Perspective on Benchmarks}.
\newblock In \emph{ACM SIGIR Forum}, volume~58, pp.\  1--27. ACM New York, NY, USA, 2025.

\bibitem[Anderljung et~al.(2023)Anderljung, Barnhart, Korinek, Leung, O'Keefe, Whittlestone, Avin, Brundage, Bullock, Cass-Beggs, et~al.]{anderljung2023frontier}
Anderljung, M., Barnhart, J., Korinek, A., Leung, J., O'Keefe, C., Whittlestone, J., Avin, S., Brundage, M., Bullock, J., Cass-Beggs, D., et~al.
\newblock Frontier {AI} regulation: Managing emerging risks to public safety.
\newblock \emph{arXiv preprint arXiv:2307.03718}, 2023.

\bibitem[Anwar et~al.(2024)Anwar, Saparov, Rando, Paleka, Turpin, Hase, Lubana, Jenner, Casper, Sourbut, et~al.]{anwar2024foundational}
Anwar, U., Saparov, A., Rando, J., Paleka, D., Turpin, M., Hase, P., Lubana, E.~S., Jenner, E., Casper, S., Sourbut, O., et~al.
\newblock Foundational challenges in assuring alignment and safety of large language models.
\newblock \emph{arXiv preprint arXiv:2404.09932}, 2024.

\bibitem[Barbosa-Silva et~al.(2022)Barbosa-Silva, Ott, Blagec, Brauner, and Samwald]{barbosa2022mapping}
Barbosa-Silva, A., Ott, S., Blagec, K., Brauner, J., and Samwald, M.
\newblock Mapping global dynamics of benchmark creation and saturation in artificial intelligence.
\newblock \emph{arXiv preprint arXiv:2203.04592}, 2022.

\bibitem[Barbour et~al.(2009)Barbour, Kleinert, Wager, and Yentis]{barbourGuidelinesRetractingArticles2009}
Barbour, V., Kleinert, S., Wager, E., and Yentis, S.
\newblock {Guidelines for Retracting Articles}.
\newblock Technical report, Committee on Publication Ethics, September 2009.

\bibitem[Barrett et~al.(2024)Barrett, Jackson, Murphy, Madkour, and Newman]{barrett2024benchmark}
Barrett, A.~M., Jackson, K., Murphy, E.~R., Madkour, N., and Newman, J.
\newblock Benchmark early and red team often: A framework for assessing and managing dual-use hazards of {AI} foundation models.
\newblock \emph{arXiv preprint arXiv:2405.10986}, 2024.

\bibitem[Beyer et~al.(2019)Beyer, L{\"o}we, and Wendler]{beyer2019reliable}
Beyer, D., L{\"o}we, S., and Wendler, P.
\newblock Reliable benchmarking: requirements and solutions.
\newblock \emph{International Journal on Software Tools for Technology Transfer}, 21\penalty0 (1):\penalty0 1--29, 2019.

\bibitem[Beyer et~al.(2020)Beyer, H{\'e}naff, Kolesnikov, Zhai, and Oord]{beyer2020we}
Beyer, L., H{\'e}naff, O.~J., Kolesnikov, A., Zhai, X., and Oord, A. v.~d.
\newblock Are we done with {ImageNet}?
\newblock \emph{arXiv preprint arXiv:2006.07159}, 2020.

\bibitem[Blagec et~al.(2023)Blagec, Kraiger, Fr{\"u}hwirt, and Samwald]{blagec2023benchmark}
Blagec, K., Kraiger, J., Fr{\"u}hwirt, W., and Samwald, M.
\newblock Benchmark datasets driving artificial intelligence development fail to capture the needs of medical professionals.
\newblock \emph{Journal of Biomedical Informatics}, 137:\penalty0 104274, 2023.

\bibitem[Campolo(2025)]{campoloStateoftheartTemporalOrder2025}
Campolo, A.
\newblock State-of-the-art: The temporal order of benchmarking culture.
\newblock 4\penalty0 (2):\penalty0 35, 2025.
\newblock ISSN 2731-4650, 2731-4669.
\newblock \doi{10.1007/s44206-025-00190-x}.
\newblock URL \url{https://link.springer.com/10.1007/s44206-025-00190-x}.

\bibitem[Chen et~al.(2021)Chen, Tworek, Jun, Yuan, Pinto, Kaplan, Edwards, Burda, Joseph, Brockman, et~al.]{chen2021evaluating}
Chen, M., Tworek, J., Jun, H., Yuan, Q., Pinto, H. P. D.~O., Kaplan, J., Edwards, H., Burda, Y., Joseph, N., Brockman, G., et~al.
\newblock Evaluating large language models trained on code.
\newblock \emph{arXiv preprint arXiv:2107.03374}, 2021.

\bibitem[Chowdhury et~al.(2024)Chowdhury, Aung, Chan, Jaffe, Sherburn, Starace, Mays, Dias, Aljubeh, Glaese, Jimenez, Yang, Ho, Patwardhan, Liu, and Madry]{chowdhury2024swebench}
Chowdhury, N., Aung, J., Chan, J.~S., Jaffe, O., Sherburn, D., Starace, G., Mays, E., Dias, R., Aljubeh, M., Glaese, M., Jimenez, C.~E., Yang, J., Ho, L., Patwardhan, T., Liu, K., and Madry, A.
\newblock {Introducing SWE-bench Verified}.
\newblock \url{https://openai.com/index/introducing-swe-bench-verified/}, August 2024.
\newblock OpenAI blog post, accessed 12 May 2025.

\bibitem[Cobbe et~al.(2021)Cobbe, Kosaraju, Bavarian, Chen, Jun, Kaiser, Plappert, Tworek, Hilton, Nakano, et~al.]{cobbe2021training}
Cobbe, K., Kosaraju, V., Bavarian, M., Chen, M., Jun, H., Kaiser, L., Plappert, M., Tworek, J., Hilton, J., Nakano, R., et~al.
\newblock Training verifiers to solve math word problems.
\newblock \emph{arXiv preprint arXiv:2110.14168}, 2021.

\bibitem[Davis(2023)]{davis2023benchmarks}
Davis, E.
\newblock Benchmarks for automated commonsense reasoning: {A} survey.
\newblock \emph{ACM Computing Surveys}, 56\penalty0 (4):\penalty0 1--41, 2023.

\bibitem[Dehghani et~al.(2021)Dehghani, Tay, Gritsenko, Zhao, Houlsby, Diaz, Metzler, and Vinyals]{dehghani2021benchmark}
Dehghani, M., Tay, Y., Gritsenko, A.~A., Zhao, Z., Houlsby, N., Diaz, F., Metzler, D., and Vinyals, O.
\newblock The benchmark lottery.
\newblock \emph{arXiv preprint arXiv:2107.07002}, 2021.

\bibitem[Eriksson et~al.(2025)Eriksson, Purificato, Noroozian, Vinagre, Chaslot, Gomez, and Fernandez-Llorca]{eriksson2025can}
Eriksson, M., Purificato, E., Noroozian, A., Vinagre, J., Chaslot, G., Gomez, E., and Fernandez-Llorca, D.
\newblock {Can We Trust AI Benchmarks? An Interdisciplinary Review of Current Issues in AI Evaluation}.
\newblock \emph{arXiv preprint arXiv:2502.06559}, 2025.

\bibitem[European~Parliament(2024)]{aiact}
European~Parliament, C. o. t. E.~U.
\newblock Regulation ({EU}) 2024/1689 of the {European Parliament and of the Council of 13 June 2024}.
\newblock \url{https://eur-lex.europa.eu/legal-content/EN/TXT/?uri=CELEX:32024R1689}, 2024.
\newblock [Accessed 26-09-2024].

\bibitem[Gebru et~al.(2021)Gebru, Morgenstern, Vecchione, Vaughan, Wallach, Iii, and Crawford]{gebruDatasheetsDatasets2021}
Gebru, T., Morgenstern, J., Vecchione, B., Vaughan, J.~W., Wallach, H., Iii, H.~D., and Crawford, K.
\newblock {Datasheets for Datasets}, December 2021.

\bibitem[Gema et~al.(2024)Gema, Leang, Hong, Devoto, Mancino, Saxena, He, Zhao, Du, Madani, et~al.]{gema2024we}
Gema, A.~P., Leang, J. O.~J., Hong, G., Devoto, A., Mancino, A. C.~M., Saxena, R., He, X., Zhao, Y., Du, X., Madani, M. R.~G., et~al.
\newblock {Are We Done with MMLU?}
\newblock \emph{arXiv preprint arXiv:2406.04127}, 2024.

\bibitem[Ghosh et~al.(2025)Ghosh, Frase, Williams, Luger, R{\"o}ttger, Barez, McGregor, Fricklas, Kumar, Bollacker, et~al.]{ghosh2025ailuminate}
Ghosh, S., Frase, H., Williams, A., Luger, S., R{\"o}ttger, P., Barez, F., McGregor, S., Fricklas, K., Kumar, M., Bollacker, K., et~al.
\newblock {AILuminate: Introducing v1. 0 of the AI Risk and Reliability Benchmark from MLCommons}.
\newblock \emph{arXiv preprint arXiv:2503.05731}, 2025.

\bibitem[Hendrycks et~al.(2020)Hendrycks, Burns, Basart, Zou, Mazeika, Song, and Steinhardt]{hendrycks2020measuring}
Hendrycks, D., Burns, C., Basart, S., Zou, A., Mazeika, M., Song, D., and Steinhardt, J.
\newblock Measuring massive multitask language understanding.
\newblock \emph{arXiv preprint arXiv:2009.03300}, 2020.

\bibitem[Kamradt(2024)]{kamradt2024needle}
Kamradt, G.
\newblock Llmtest\_needleinahaystack, 2024.
\newblock URL \url{https://github.com/gkamradt/LLMTest_NeedleInAHaystack}.
\newblock Accessed: 2025-05-12.

\bibitem[Kutuzov et~al.(2018)Kutuzov, {\O}vrelid, Szymanski, and Velldal]{kutuzov2018diachronic}
Kutuzov, A., {\O}vrelid, L., Szymanski, T., and Velldal, E.
\newblock Diachronic word embeddings and semantic shifts: a survey.
\newblock \emph{arXiv preprint arXiv:1806.03537}, 2018.

\bibitem[{LangChain}(2024)]{langchain2024multi}
{LangChain}.
\newblock Multi needle in a haystack, March 2024.
\newblock URL \url{https://blog.langchain.dev/multi-needle-in-a-haystack/}.
\newblock Accessed: 2025-05-12.

\bibitem[Liao et~al.(2021)Liao, Taori, Raji, and Schmidt]{liao2021we}
Liao, T., Taori, R., Raji, I.~D., and Schmidt, L.
\newblock Are we learning yet? {A} meta review of evaluation failures across machine learning.
\newblock In \emph{Thirty-fifth Conference on Neural Information Processing Systems Datasets and Benchmarks Track (Round 2)}, 2021.

\bibitem[Linghan et~al.(2024)Linghan, Jianjun, Ying, Jingwu, Xuzhi, Zhifeng, and Xiaoben]{linghan2024chinagov}
Linghan, Z., Jianjun, Y., Ying, C., Jingwu, Z., Xuzhi, H., Zhifeng, Z., and Xiaoben, X.
\newblock {Artificial Intelligence Law of the People’s Republic of China (Draft for Suggestions from Scholars)}.
\newblock \emph{Center for Security and Emerging Technology. Accessed: Aug}, 16, 2024.

\bibitem[Luccioni et~al.(2022)Luccioni, Corry, Sridharan, Ananny, Schultz, and Crawford]{luccioni2022framework}
Luccioni, A.~S., Corry, F., Sridharan, H., Ananny, M., Schultz, J., and Crawford, K.
\newblock {A framework for deprecating datasets: Standardizing documentation, identification, and communication}.
\newblock In \emph{Proceedings of the 2022 acm conference on fairness, accountability, and transparency}, pp.\  199--212, 2022.

\bibitem[Maslej et~al.(2025)Maslej, Fattorini, Perrault, Gil, Parli, Kariuki, Capstick, Reuel, Brynjolfsson, Etchemendy, Ligett, Lyons, Manyika, Niebles, Shoham, Wald, Walsh, Hamrah, Santarlasci, Lotufo, Rome, Shi, and Oak]{maslej2025aiindex}
Maslej, N., Fattorini, L., Perrault, R., Gil, Y., Parli, V., Kariuki, N., Capstick, E., Reuel, A., Brynjolfsson, E., Etchemendy, J., Ligett, K., Lyons, T., Manyika, J., Niebles, J.~C., Shoham, Y., Wald, R., Walsh, T., Hamrah, A., Santarlasci, L., Lotufo, J.~B., Rome, A., Shi, A., and Oak, S.
\newblock Artificial intelligence index report 2025, 2025.
\newblock URL \url{https://arxiv.org/abs/2504.07139}.

\bibitem[McIntosh et~al.(2024)McIntosh, Susnjak, Arachchilage, Liu, Watters, and Halgamuge]{mcintosh2024inadequacies}
McIntosh, T.~R., Susnjak, T., Arachchilage, N., Liu, T., Watters, P., and Halgamuge, M.~N.
\newblock Inadequacies of large language model benchmarks in the era of generative artificial intelligence.
\newblock \emph{arXiv preprint arXiv:2402.09880}, 2024.

\bibitem[Miltenberger et~al.(2023)Miltenberger, Arzt, Holzinger, and N{\"a}umann]{miltenberger2023benchmarking}
Miltenberger, M., Arzt, S., Holzinger, P., and N{\"a}umann, J.
\newblock {Benchmarking the benchmarks}.
\newblock In \emph{Proceedings of the 2023 ACM Asia Conference on Computer and Communications Security}, pp.\  387--400, 2023.

\bibitem[Paskin(2010)]{paskin2010digital}
Paskin, N.
\newblock Digital object identifier (doi{\textregistered}) system.
\newblock \emph{Encyclopedia of library and information sciences}, 3:\penalty0 1586--1592, 2010.

\bibitem[Polo et~al.(2024)Polo, Weber, Choshen, Sun, Xu, and Yurochkin]{polo2024tinybenchmarks}
Polo, F.~M., Weber, L., Choshen, L., Sun, Y., Xu, G., and Yurochkin, M.
\newblock {tinyBenchmarks: evaluating LLMs with fewer examples}.
\newblock \emph{arXiv preprint arXiv:2402.14992}, 2024.

\bibitem[Raji et~al.(2021{\natexlab{a}})Raji, Denton, Bender, Hanna, and Paullada]{rajiAIEverythingWhole2021}
Raji, D., Denton, E., Bender, E.~M., Hanna, A., and Paullada, A.
\newblock {{AI}} and the {Everything in the Whole Wide World Benchmark}.
\newblock In Vanschoren, J. and Yeung, S. (eds.), \emph{Proceedings of the {{Neural Information Processing Systems Track}} on {{Datasets}} and {{Benchmarks}}}, volume~1, 2021{\natexlab{a}}.

\bibitem[Raji et~al.(2021{\natexlab{b}})Raji, Bender, Paullada, Denton, and Hanna]{raji2021ai}
Raji, I.~D., Bender, E.~M., Paullada, A., Denton, E., and Hanna, A.
\newblock Ai and the everything in the whole wide world benchmark.
\newblock \emph{arXiv preprint arXiv:2111.15366}, 2021{\natexlab{b}}.

\bibitem[Rauh et~al.(2024)Rauh, Marchal, Manzini, Hendricks, Comanescu, Akbulut, Stepleton, Mateos-Garcia, Bergman, Kay, Griffin, Bariach, Gabriel, Rieser, Isaac, and Weidinger]{rauh2024gapsinsafetyeval}
Rauh, M., Marchal, N., Manzini, A., Hendricks, L.~A., Comanescu, R., Akbulut, C., Stepleton, T., Mateos-Garcia, J., Bergman, S., Kay, J., Griffin, C., Bariach, B., Gabriel, I., Rieser, V., Isaac, W., and Weidinger, L.
\newblock {Gaps in the Safety Evaluation of Generative AI}.
\newblock \emph{Proceedings of the AAAI/ACM Conference on AI, Ethics, and Society}, 7\penalty0 (1):\penalty0 1200--1217, Oct. 2024.
\newblock \doi{10.1609/aies.v7i1.31717}.
\newblock URL \url{https://ojs.aaai.org/index.php/AIES/article/view/31717}.

\bibitem[Ravaut et~al.(2024)Ravaut, Ding, Jiao, Chen, Li, Zhao, Qin, Xiong, and Joty]{ravaut2024much}
Ravaut, M., Ding, B., Jiao, F., Chen, H., Li, X., Zhao, R., Qin, C., Xiong, C., and Joty, S.
\newblock How much are {LLM}s contaminated? {A} comprehensive survey and the llmsanitize library.
\newblock \emph{arXiv preprint arXiv:2404.00699}, 2024.

\bibitem[Ren et~al.(2024)Ren, Basart, Khoja, Gatti, Phan, Yin, Mazeika, Pan, Mukobi, Kim, et~al.]{ren2024safetywashing}
Ren, R., Basart, S., Khoja, A., Gatti, A., Phan, L., Yin, X., Mazeika, M., Pan, A., Mukobi, G., Kim, R., et~al.
\newblock {Safetywashing: Do AI Safety Benchmarks Actually Measure Safety Progress?}
\newblock \emph{Advances in Neural Information Processing Systems}, 37:\penalty0 68559--68594, 2024.

\bibitem[Reuel et~al.(2024)Reuel, Hardy, Smith, Lamparth, Hardy, and Kochenderfer]{reuel2024betterbenchassessingaibenchmarks}
Reuel, A., Hardy, A., Smith, C., Lamparth, M., Hardy, M., and Kochenderfer, M.~J.
\newblock {BetterBench: Assessing AI Benchmarks, Uncovering Issues, and Establishing Best Practices}, 2024.
\newblock URL \url{https://arxiv.org/abs/2411.12990}.

\bibitem[Robison(2025)]{robisonMetaGotCaught2025}
Robison, K.
\newblock Meta got caught gaming {{AI}} benchmarks.
\newblock 2025.
\newblock URL \url{https://www.theverge.com/meta/645012/meta-llama-4-maverick-benchmarks-gaming}.

\bibitem[Singh et~al.(2025)Singh, Nan, Wang, D'Souza, Kapoor, {\"U}st{\"u}n, Koyejo, Deng, Longpre, Smith, et~al.]{singh2025leaderboard}
Singh, S., Nan, Y., Wang, A., D'Souza, D., Kapoor, S., {\"U}st{\"u}n, A., Koyejo, S., Deng, Y., Longpre, S., Smith, N., et~al.
\newblock {The Leaderboard Illusion}.
\newblock \emph{arXiv preprint arXiv:2504.20879}, 2025.

\bibitem[Srivastava et~al.(2022)Srivastava, Rastogi, Rao, Shoeb, Abid, Fisch, Brown, Santoro, Gupta, Garriga-Alonso, et~al.]{srivastava2022beyond}
Srivastava, A., Rastogi, A., Rao, A., Shoeb, A. A.~M., Abid, A., Fisch, A., Brown, A.~R., Santoro, A., Gupta, A., Garriga-Alonso, A., et~al.
\newblock Beyond the imitation game: Quantifying and extrapolating the capabilities of language models.
\newblock \emph{arXiv preprint arXiv:2206.04615}, 2022.

\bibitem[Staufer et~al.(2025)Staufer, Yang, Reuel, and Casper]{stauferAuditCardsContextualizing2025}
Staufer, L., Yang, M., Reuel, A., and Casper, S.
\newblock {Audit Cards: Contextualizing {{AI}} Evaluations}.
\newblock \emph{arXiv preprint arXiv:2504.13839}, April 2025.

\bibitem[Varoquaux et~al.(2024)Varoquaux, Luccioni, and Whittaker]{varoquaux2024hype}
Varoquaux, G., Luccioni, A.~S., and Whittaker, M.
\newblock {Hype, Sustainability, and the Price of the Bigger-is-Better Paradigm in AI}.
\newblock \emph{arXiv preprint arXiv:2409.14160}, 2024.

\bibitem[Xia et~al.(2024)Xia, Deng, Dunn, and Zhang]{xia2024agentlessdemystifyingllmbasedsoftware}
Xia, C.~S., Deng, Y., Dunn, S., and Zhang, L.
\newblock {Agentless: Demystifying LLM-based Software Engineering Agents}, 2024.
\newblock URL \url{https://arxiv.org/abs/2407.01489}.

\bibitem[Xu et~al.(2024{\natexlab{a}})Xu, Guan, Greene, Kechadi, et~al.]{xu2024benchmark}
Xu, C., Guan, S., Greene, D., Kechadi, M., et~al.
\newblock {Benchmark Data Contamination of Large Language Models: A Survey}.
\newblock \emph{arXiv preprint arXiv:2406.04244}, 2024{\natexlab{a}}.

\bibitem[Xu et~al.(2024{\natexlab{b}})Xu, Wang, Fan, and Liu]{xu2024benchmarking}
Xu, R., Wang, Z., Fan, R.-Z., and Liu, P.
\newblock Benchmarking benchmark leakage in large language models.
\newblock \emph{arXiv preprint arXiv:2404.18824}, 2024{\natexlab{b}}.

\bibitem[Zhou et~al.(2023)Zhou, Zhu, Chen, Chen, Zhao, Chen, Lin, Wen, and Han]{zhou2023don}
Zhou, K., Zhu, Y., Chen, Z., Chen, W., Zhao, W.~X., Chen, X., Lin, Y., Wen, J.-R., and Han, J.
\newblock Don't make your {LLM} an evaluation benchmark cheater.
\newblock \emph{arXiv preprint arXiv:2311.01964}, 2023.

\end{thebibliography}
\bibliographystyle{icml2025}

\newpage
\appendix
\onecolumn
\section{Sample Scenarios of the Deprecation Workflow}
\label{appendix:cases}
For these illustrative examples, we suppose that the benchmark developer is a research group affiliated with a university (Group), while the governance actor is a safety institute (Institute). The benchmark measures a model's capabilities to assist in synthesizing controlled chemical substances, relevant to the safety evaluation of frontier models. Several scenarios may  occur:

\begin{enumerate}
    \item \emph{The Group deprecates the benchmark, while the Institute tracks and records the deprecation.} The Group, who developed the benchmark, discovers problems with the benchmark and decides to initiate the deprecation process themselves. The Group creates the deprecation report. The safety institute simply records the deprecated benchmark alongside its deprecation report in their deprecation list.

    \item \emph{The Group does not deprecate the benchmark, even when requested by the Institute. The Institute deprecates the benchmark.} The Group ran out of funding to maintain the benchmark a few months before receiving the deprecation request from the Agency. Therefore, there is nobody to maintain it. After a pre-defined period waiting for a reply, the Agency decides to include the benchmark in their deprecation list and creates the deprecation report.

    \item \emph{The Institute deprecates the benchmark, and only notifies the Group later.} The Institute includes the benchmark in their deprecation list and creates the deprecation report. In the report, alongside explaining the flaws found in the benchmark, the Institute reasons that the inclusion of the benchmark in their list was a matter of urgency to maintain appropriate monitoring for chemical synthesis capabilities of select frontier models. The Group appeals, mentioning that they can correct the flaws. The Agency responds to the appeal after a pre-defined timeline by saying that the benchmark should incorporate more data sources and incorporate new synthetic chemical compounds of interest. The Group then presents an upgraded version of the benchmark following the Agency's recommendations. The Group notifies the Institute about the upgraded version and the Institute updates its deprecation report to direct readers to use the upgraded version.
\end{enumerate}

\clearpage
\section{Sample Deprecation Reports}
\label{appendix:deprecation_report}
These examples are based on real problems discovered in the benchmarks, while the deprecation process presented here is purely hypothetical. They are intended to demonstrate how deprecation reports can transparently document scope, rationale, implications for historical results, and pathways for transition.

\begin{table*}[htb]
    \centering
    \mbox{}\clap{
        \setlength{\tabcolsep}{0.8em} 
        \renewcommand{\arraystretch}{1.2}
        \begin{tabular}{|p{3.5cm}|p{6.5cm}|p{6.5cm}|}
            \hline
            \textbf{} & \textbf{Partial Deprecation: SWE-bench (v1.0)}& \textbf{Partial Deprecation: SWE-bench Lite (v1.0)} \\
            \hline
            \textbf{Issuing Authority} & [governance institution] \newline \texttt{deprecation@institution.gov} & [benchmark developers] \newline \texttt{contact@swebench.com} \\
            \hline
            \textbf{Benchmark Version} & SWE-bench v1.0 & SWE-bench Lite v1.0 \\
            \hline
            \textbf{Benchmark Link} & \url{https://github.com/SWE-bench/SWE-bench} & \url{https://huggingface.co/datasets/princeton-nlp/SWE-bench_Lite}\\
            \hline
            \textbf{Benchmark Developer} & Jimenez et al. \newline Princeton NLP Group& Jimenez et al. \newline Princeton NLP Group\\
            \hline
            \textbf{Affected Components} & Subset mentioned below& Subset mentioned below\\
            \hline
            \textbf{Justification} & ``38.3\% of samples were flagged for underspecified problem statements, and 61.1\% were flagged for unit tests that may unfairly mark valid solutions as incorrect. Overall, our annotation process resulted in 68.3\% of SWE-bench samples being filtered out due to underspecification, unfair unit tests, or other issues" \citep{chowdhury2024swebench}& ``SWE-bench Lite contains problems with exact ground truth patch in the description (4.3\%), problems with missing critical information needed to solve the issue (10.0\%), and problems that include misleading solutions in the issue description (5.0\%)" \citep{xia2024agentlessdemystifyingllmbasedsoftware}\\
            \hline
            \textbf{Deprecation Timeline} & \textbf{Announcement:} May 2025 \newline \textbf{Deprecation Effective:} November 2025 & \textbf{Announcement:} May 2025 \newline \textbf{Deprecation Effective:} August 2025 \\
            \hline
            \textbf{Alternatives} & SWE-Bench Verified \citep{chowdhury2024swebench}& SWE-Bench-Lite-S, a filtered subset. \newline \url{https://github.com/OpenAutoCoder/Agentless/tree/main/classification}\\
            \hline
            \textbf{Changes in Benchmark Usage} & Recommended to switch to alternative, which is generally improved. For deprecated tasks: \newline - Exclude from future evaluations \newline - Do not use for model comparisons \newline For remaining tasks: \newline - Continue usage with caution \newline - Clearly document any limitations& Recommended to switch to filtered variant. For deprecated tasks: \newline - Exclude from future evaluations \newline - Do not use for model comparisons \newline For remaining tasks: \newline - Continue usage with caution \newline - Clearly document any limitations\\
            \hline
            \textbf{Guidance on Historical Results} & Historical results from SWE-Bench v1.0 are no longer valid for comparison due to identified flaws. Users should transition to SWE-Bench Verified for future evaluations. & Historical results should be re-evaluated: \newline - Exclude deprecated tasks from analyses \newline - Reassess model performance using SWE-Bench-Lite-S \\
            \hline
        \end{tabular}
    }
    
    \caption{Sample deprecation reports for SWE-bench Benchmarks. Note that deprecation details are fictional, while the mentioned flaws in the benchmarks are real.}
\end{table*}

\end{document}